\documentclass[preprint]{aastex}

\begin{document}

\title{Mutual Constraints Between Reionization Models and Parameter Extraction From Cosmic Microwave Background Data} 
\author{Aparna Venkatesan} 
\affil{CASA, Department of Astrophysical and Planetary Sciences, \\
University of Colorado, 389 UCB, Boulder, CO 80309-0389}
\email{aparna@casa.colorado.edu}

\begin{abstract}

Spectroscopic studies of high-redshift objects and increasingly precise
data on the cosmic microwave background (CMB) are beginning to
independently place strong complementary bounds on the epoch of hydrogen
reionization. Parameter estimation from current CMB data continues,
however, to be subject to several degeneracies. Here, we focus on those
degeneracies in CMB parameter forecasts related to the optical depth to
reionization. We extend earlier work on the mutual constraints that such
analyses of CMB data and a reionization model may place on each other to a
more general parameter set, and to the case of data anticipated from the
$MAP$ satellite. We focus in particular on a semi-analytic model of
reionization by the first stars, although the methods here are easily
extended to other reionization scenarios. A reionization model can provide
useful complementary information for cosmological parameter extraction from
the CMB, particularly for the degeneracies between the optical depth and
either of the amplitude and index of the primordial scalar power spectrum,
which are still present in the most recent data. Alternatively, by using a
reionization model, known limits on astrophysical quantities can reduce the
forecasted errors on cosmological parameters.  Forthcoming CMB data also
have the potential to constrain the sites of early star formation, as well
as the fraction of baryons that participate in it, if reionization were
caused by stellar activity at high redshifts.  Finally, we examine the
implications of an independent, e.g., spectroscopic, determination of the
epoch of reionization for the determination of cosmological parameters from
the CMB. This has the potential to significantly strengthen limits from the
CMB on parameters such as the index of the power spectrum, while having the
considerable advantage of being free of the choice of the reionization
model.

\end{abstract}


\section{Introduction}

The rapid progress in detector technology has led to the successful
operation of many ground- and balloon-based experiments in the last few
years for measuring the anisotropies in the CMB. Analyses of the recent
data from experiments such as Boomerang \citep{boom}, MAXIMA-1
\citep{maxima}, and DASI \citep{dasi} have confirmed the adiabatic cold
dark matter (CDM) paradigm for describing the development of structure and
the properties of the power spectrum of the CMB. They have also revealed
that the universe is close to being spatially flat, and have begun to place
tight constraints, in advance of satellite CMB experiments, on the
cosmological parameters that describe our universe.  Analyses of present
data (see papers above, and those of, e.g., \citet{tzh01} and
\citet{wangtz01}) indicate, however, that strong degeneracies are still
present in parameter extraction from the CMB, so that techniques to break
these degeneracies continue to be valuable at present. Many of these
degeneracies had been anticipated on theoretical grounds, and several
methods to break them using observations of Type Ia SNe \citep{efst99},
weak lensing \citep{hulens}, redshift surveys \citep{eht99, wang99,
popa01}, or combinations of these \citep{efstbond} have been proposed.
Ongoing and future CMB observations\footnote{Compilations of and links to
various CMB experiments may be found at:
http://www.hep.upenn.edu/$\sim$max/cmb/experiments.html, and
http://background.uchicago.edu/$\sim$whu/cmbex.html} should provide
markedly improved constraints on degenerate parameters through the
detection of polarization in the CMB at large angular scales
\citep{staggs}, and through dramatically increased sky coverage in the case
of satellite experiments such as $MAP$\footnote{http://map.gsfc.nasa.gov .}
or $Planck$\footnote{http://astro.estec.esa.nl/SA-general/Projects/Planck
.}. The latter is especially important for overcoming cosmic variance for
CMB multipoles, $l \la $ 100. Current CMB data on the temperature
anisotropy at degree and sub-degree scales provide an upper limit of about
0.3 for the optical depth to reionization, which may be translated to a
model-dependent constraint on the redshift of hydrogen reionization,
$z_{\rm reion}$ $\la$ 25 \citep{wangtz01}.

Spectroscopic studies of high-$z$ quasars and galaxies blueward of
Ly$\alpha$ have revealed the lack of a H~I Gunn-Peterson (GP) trough,
implying that the intergalactic medium (IGM) is highly ionized up to $z
\sim 6$ \citep{fan1,dey,hugal}. Recently, \citet{djor} presented
observations of quasars at $ z \ga 5.2$ indicating a steady increase in the
opacity of the Ly$\alpha$ forest for $z \sim$ 5.2--5.7, while
\citet{becker} presented a detection of the GP trough in the spectrum of
the highest-redshift quasar known to date at $z \sim 6.3$
\citep{fan3}. Together, these data may be an indication of the epoch of H~I
reionization occurring not far beyond $z \sim 6$. As these authors have
taken care to note, the detection of the GP trough in a single
line-of-sight is not definitive evidence of $z_{\rm reion} \sim 6$; it may,
however, be probing the end of the gradual process of inhomogeneous
reionization coinciding with the disappearance of the last neutral regions
in the high-$z$ IGM. This would be consistent with the lower end of the
range of redshifts, $z \sim$ 6--20, predicted by theoretical models for H~I
reionization, either semi-analytic
\citep{tsb94,gs96,hl97,valsilk,mhrees99,mirees00} or based on numerical
simulations \citep{cenost93,gnedin00,ciardi,benson}. Recent reviews of
reionization may be found in \citet{shapiro}, and \citet{loebbark}.

The reionization of the IGM subsequent to recombination at $z \sim$ 1000 is
thought to have been caused by increasing numbers of the first luminous
sources. Although a variety of astrophysical objects or processes could
have reionized the IGM, most of these group into stellar-related or
QSO-related models, or equivalently, into sources with soft (star-like) or
hard (QSO-like) ionizing spectra\footnote{We note that this division of
source populations according to their spectral properties will no longer be
valid if the first stars generated hard ionizing radiation, e.g., if they
formed in an initial mass function biased towards extremely high masses, or
if reionization were caused principally by metal-free stars
\citep{tumshull}.}. Of these models, photoionization by stars and
``mini-quasars'' are at present the leading scenarios
\citep{hk99,loebbark}, with the large majority of currently accepted
reionization models involving stellar-type radiation for the following
observationally motivated reasons.  First, the space density of the large,
optically bright QSO population appears to decrease after a peak at $z
\sim$ 3. This has been confirmed by optical observations up to $z \sim$ 6.3
\citep{fan3}, and corroborated by radio surveys \citep{shaver} which should
not suffer from the effects of dust obscuration.  QSOs may however still be
relevant to reionization, if they are powered by massive black holes that
are postulated to form as a fixed universal fraction of the mass of
collapsing halos at all redshifts \citep{hl98,valsilk}. This leads to a
large population of faint QSOs (mini-quasars) in small halos at $z \ga 6$
that are currently undetected.  The observed turnover in the QSO space
density at $z \ga 3$ would then be true only for the brightest QSOs; this
population, however, appears unlikely to cause H~I reionization, either
through their UV photons (\citet{gs96}, and references therein;
\citet{fan3}) or through the associated X-rays \citep{venkgs01}.  Second,
if QSOs (mini- or otherwise) reionized the universe, we would expect the
H~I and He~II reionization epochs to be coeval, given that QSOs are copious
producers of H~I and He~II ionizing photons.  The current data indicate
that this does not occur, with He~II reionization occurring at $z \sim$ 3
\citep{kriss}, and that of H~I reionization before $z \sim 6$. Thus the
delayed reionization of He~II relative to that of H~I would seem to imply a
metagalactic ionizing background dominated by a soft spectrum.

One might counter these two reasons with the argument that stars and
quasars have similar effective ionizing power, which has been made
frequently, most recently by \citet{barkana1}, and which goes as follows.
The average efficiency with which the baryons in a high-$z$ halo form black
holes that could power mini-QSOs is likely less than that for star
formation.  However, this is balanced by the higher escape fraction of
ionizing radiation from mini-QSOs, given their inherently harder spectrum,
leading to roughly the same overall output of IGM-ionizing photons per
baryon in luminous objects. We note here that such arguments are limited by
their not considering the detailed source spectrum which directly
influences the growth of H~II and He~III regions, so that the observed
delay in the reionization epochs of H~I and He~II remains unresolved
quantitatively. Lastly, stars can account for the ubiquitous trace
metallicity of about 0.003 of solar values seen in the low-density
Ly$\alpha$ forest clouds up to $z \sim$ 4 (\citet{venk}, and references
therein). A simple calculation shows that this detected metallicity implies
a minimum (on average) of ten stellar ionizing photons per baryon having
been generated in the past. This again implies that the first stars must
play some role in reionization.

In summary, while the nature of the reionizing sources are at present
unknown, the data suggest that their spectral properties resemble those of
stars rather than quasars, and that radiation from the first stars is
likely to have played a significant, if not the dominant, role in H~I
reionization. Therefore, in this work we will focus on a stellar origin for
the reionizing spectrum, although we emphasize that one cannot at present
rule out the possibility of alternate sources or a combination of high-$z$
source populations with individually varying spectral hardness that cause
reionization.  We refer the reader to \citet{gs96} for an excellent
discussion on the relative roles of reionizing sources whose spectra are
star-like, QSO-like and of intermediate spectral hardness. From this point
onwards, reionization is always meant to refer to that of H~I, rather than
He~II.

In this paper, we focus on those degeneracies in CMB parameter forecasts
that involve the optical depth to reionization, $\tau$, based on methods
developed in a previous work (\citet{venk}; henceforth Paper I) that
examined the valuable complementary information provided by a reionization
model. Typically, in CMB parameter extraction, the universe is assumed to
reionize abruptly, leading to discretized values of $\tau$ in the
multi-dimensional grid of models being tested in likelihood analyses of the
data. This does not utilize, however, the strong sensitivity of $z_{\rm
reion}$, and hence $\tau$, to specific parameters such as the spectral
index of the primordial scalar power spectrum. As we noted in Paper I,
$\tau$ is unique by definition amongst the set of standard cosmological
parameters extracted from CMB data, being the only quantity which is not
determined purely by the physics prior to the first few minutes after the
Big Bang. Thus, it can potentially provide information on
post-recombination astrophysical processes, if the other (cosmological)
parameters which affect $\tau$ are well-constrained.  We extend Paper I
here to a larger parameter set in a $\Lambda$CDM cosmology; in the spirit
of timeliness, we specifically consider the constraints anticipated from
the data from the recently launched $MAP$ satellite, and we also include in
our analysis the implications of an independent, e.g., spectroscopic,
determination of $z_{\rm reion}$. Other improvements are detailed in the
next section.

The plan of this paper is as follows. In \S 2, we review the reionization
model that we consider and the formalism related to CMB parameter
estimation. In \S 3, we present our results on the projected parameter
yield from $MAP$, and we detail how a reionization model may improve
constraints on cosmological parameters determined from the CMB, and vice
versa. In \S 4, we discuss the implications of the secondary anisotropies
generated in the CMB during reionization for the analysis in this work, and
summarize the observations that are likely to best constrain the various
aspects of reionization in the future.  We conclude in \S 5.

\section{Overview of the Reionization Model and CMB Analysis}

The analysis in this paper essentially follows the methods developed in
Paper I, which is extended here for a $\Lambda$CDM model; the points of
departure and improvements here are described below.

We assume that stars are responsible for reionization (for the reasons
presented in \S 1), and use the semi-analytic stellar reionization model
developed by \citet{hl97}, with the modifications described in Paper I. We
take the primordial matter power spectrum of density fluctuations to be,
$P(k) \propto k^n$ $T^2 (k)$, where $n$ is the index of the scalar power
spectrum, and the matter transfer function $T(k)$ is taken from
\citet{ehu98}. We normalize $P(k)$ to the present-day rms density contrast
over spheres of radius 8 $h^{-1}$ Mpc, $\sigma_8$.

We track the fraction of all baryons in star-forming halos, $F_{\rm B}$, by
the Press-Schechter formalism, allowing star formation only in halos of
virial temperature $\ga$ 10$^4$ K, corresponding to the mass threshold for
the onset of hydrogen line cooling. The details concerning the adopted
stellar spectrum of ionizing photons and the solution for the growth of
ionization regions around individual halos may be found in Paper I. We
define reionization as the epoch of overlap of individual H~II regions,
i.e., when the volume filling factor of ionized hydrogen, $F_{\rm H \, II}$
= 1.  We include the effects of inhomogeneity in the IGM through a clumping
factor, $c_L$, rather than assuming a smooth IGM as in Paper I. We define
$c_L$ to be the space-averaged clumping factor of ionized hydrogen, $c_L$
$\equiv$ $<$~$n_p^2$~$>$/$<n_p>^2$ \citep{sg87,mhrees99}, which is
equivalent to $<n_e^2>$/$<$~$n_e$$>^2$ in this work as the sources of
photoionization do not generate any helium-ionizing photons.  The optical
depth to reionization from electron scattering is then given by:

\begin{equation}
\tau  \simeq  0.057 \; \Omega_{\rm b} h \; \int_0^{z_{\rm reion}} dz \;
\frac{(1 + z)^2 \; [1 - f_\star F_{\rm B}(z)] \; F_{\rm H \, II}(z)}{\sqrt{\Omega_{\Lambda} + (1 + z)^2(1 - \Omega_{\Lambda} + \Omega_{\rm m} z)}} .
\end{equation}

The optical depth to reionization depends upon a number of parameters as,
$\tau$ = $f$($\sigma_8$, $\Omega_b$, $h$, $n$, $\Omega_\Lambda$,
$\Omega_{\rm M}$, $f_\star$, $f_{esc}$), where $f_\star$ is the fraction of
baryons in each galaxy halo forming stars, $f_{esc}$ is the escape fraction
of H~I ionizing photons from individual halos, $h$ is the Hubble constant
in units of 100 km s$^{-1}$ Mpc$^{-1}$, and the other symbols have their
usual meanings. We fix $\Omega_{\rm K}$ = 1 - $\Omega_{\rm M}$ -
$\Omega_\Lambda$. We also set $f_{esc}$ = 0.1 \citep{dsf,leitherer}, so
that it is no longer a free parameter as it was in Paper I, for the
following reasons. First, the mass threshold scale for the Press-Schechter
evolution of halos in our model corresponds to massive halos of virial
temperature $\ga 10^4$ K. The baryons in such halos are likely to be
collisionally ionized, at least partly, so that $f_{esc}$ $\ll$ 1 is
unlikely.  The values of $f_{esc}$ in low-mass systems (masses $\la$ $10^7
M_\odot$) at high redshift have been studied by \citet{rshull00}.  Second,
as shown in \citet{hl97} and Paper I (see in particular Table 1 and the
associated discussion), $\tau$ is not very sensitive to the chosen values
of $f_{esc}$, once they exceed a few percent. Third, limits on $\tau$ from
the CMB, being a single number, can be translated to a constraint on any
one non-cosmological parameter that determines $\tau$; recall, for example,
that in Paper I, both $f_\star$ and $f_{esc}$ could not be constrained
simultaneously from the CMB. Hence we choose to retain $f_\star$ as the
primary astrophysical input parameter, as $\tau$ is most sensitive to it in
our chosen reionization model.

To be complete, we note that observations of Lyman-continuum emission from
Lyman-break galaxies at $z \sim$ 3.4 by \citet{steidel} indicate values of
$f_{esc}$ exceeding 0.5. Also, some simulations of reionization by stars
often appear to require or imply similarly high values for $f_{esc}$
\citep{gnedin00,benson2} in order to have $z_{\rm reion}$ exceed $\sim$
7. The large derived value for $f_{esc}$ in the former case arises partly
from the definition itself of $f_{esc}$; as \citet{steidel} noted, their
chosen observational procedure normalized the escape fraction of 900\AA\
photons to that of 1500\AA\ photons.  Data from the local universe
\citep{dehar,leitherer}, especially of high-mass systems, generally do not
support values of $f_{esc}$ exceeding about 10\%.

Reionization affects the CMB through the Thomson scattering of CMB photons
from free electrons in the IGM. This leads to an overall damping of the
primary temperature and polarization anisotropies in the CMB, except at the
largest angular scales (small $l$), and the generation of a new feature in
the polarization power spectrum. The first effect can be distinguished from
CMB anisotropies with slightly lower peak amplitudes (corresponding to a
lower $\sigma_8$ in our model) only at the lowest $l$s, but cosmic variance
obscures the difference at such scales.  This is the origin of the
amplitude--reionization degeneracy in the CMB temperature power
spectrum. However, the reionized IGM creates a linear polarization signal
which peaks at the horizon size at $z_{\rm reion}$, so that the amplitude
and angular location of this new feature are comparatively direct probes of
the values of $\tau$ and $z_{\rm reion}$ respectively \citep{zal97b}. A
detection of polarization in the CMB at large angular scales can therefore
constrain $\tau$ far more accurately than can temperature data alone, and
has the potential to break the above degeneracy. In practice, it may prove
difficult to measure, given that the polarization anisotropy is expected to
be only at the $\sim$ 10\% level relative to that in the CMB's temperature,
and that for late reionization the above feature has an extremely small
amplitude (see next section). Additionally, foregrounds are likely to
complicate the extraction of a polarization signal at low $l$. As we do not
consider tensor contributions to the primordial matter power spectrum,
polarization here refers to the E-channel type. Lastly, we do not
explicitly consider the effects of any secondary anisotropies generated in
the CMB during reionization, but we return to this topic in \S 4.

Parameter extraction from the CMB is based on the methods outlined in Paper
I.  For cases involving $\tau$ and a set of cosmological parameters, we
follow the industry-tested Fisher matrix formalism in, e.g.,
\citet{knox95}, \citet{jung96}, \citet{zal97a}, and \citet{bet97}. If we
expand the angular power spectrum of the CMB in terms of its multipole
moments $C_l$, and assume Gaussian initial perturbations and that the $C_l$
are determined by a fiducial set of parameters describing the ``true''
universe, then we can quantify the behavior of the likelihood function of
observing any set of $C_l$s near its maximum, given the fiducial parameter
set, in terms of the Fisher information matrix, $F_{ij}$. If we further
assume that the likelihood function has a Gaussian form near its maximum,
the elements of $F_{ij}$ can be expressed as the product of pairs of
derivatives of the $C_l$ with respect to the appropriate parameters.  The
Fisher matrix represents the best accuracy with which parameters in the
chosen ``true'' model can be estimated from a CMB data set. The inverse of
$F_{ij}$ is the covariance matrix between the parameters; the minimum 1
$\sigma$ error in a parameter $P_i$ is given by $\sqrt{(F^{-1})_{ii}}$.

The reionization model, as described above, yields $\tau$ =
$\tau$($\sigma_8$, $\Omega_b$, $h$, $n$, $\Omega_\Lambda$, $\Omega_{\rm
M}$, $f_\star$) = $\tau$($P_{\rm cosmo}$, $f_\star$), while the CMB data
determines [$P_{\rm cosmo}, \tau (P_{\rm cosmo}, f_\star)$]. We can
therefore use a reionization model to relate and mutually constrain
($P_{\rm cosmo}, f_\star$).  In such cases, the derivatives of the CMB
multipoles, $C_l$, that are used to construct the Fisher matrix become
(Paper I):

\begin{equation}
\frac{\partial C_l}{\partial P_{\rm cosmo}} = \; \left. \frac{\partial
C_l}{\partial P_{\rm cosmo}} \right|_{\tau} \; + \; \;
\left. \frac{\partial C_l}{\partial \tau} \right|_{P_{\rm
cosmo}} \frac{\partial \tau}{\partial P_{\rm cosmo}} 
\end{equation}
\begin{equation}
\frac{\partial C_l}{\partial f_\star} = \; \frac{\partial
C_l}{\partial \tau}  \frac{\partial \tau}{\partial f_\star} 
\end{equation}

Parameter estimation is performed using theoretical CMB power spectra
generated by CMBFAST [version 4.0; \citet{selzal}].  This version of
CMBFAST corrects a bug in previous versions related to some models with
non-zero values of $\tau$, and includes an improved treatment of
recombination based on the work of \citet{seager}. Although the primordial
power spectrum in the reionization model and the CMB may be normalized to
$\sigma_8$ or the COBE normalization (both of which themselves depend on a
number of cosmological parameters), we choose the former option for the
following reasons.  First, there is a large difference between the physical
scales probed by COBE on which linear physics operates, and those that are
relevant for reionization and the formation of the first luminous objects
(tens of kpc and below).  In order to bridge this gap between mass
fluctuations in the linear and highly nonlinear regimes, it is more
consistent to use a parameter such as $\sigma_8$, which probes the
amplitude of the fluctuations in the power spectrum today at intermediate
scales. This choice is particularly important in a work such as this, where
constraints from a model that describes the activity of the reionizing
sources are combined with those from the CMB. Second, $\sigma_8$, like the
COBE normalization, is a well-defined observational parameter that is
currently measured to within 10\% error.  As reionization is sensitive,
however, to the amount of power on small scales, the uncertainty in the
value of $\sigma_8$ translates to a lower relative error in the amount of
small-scale power than does the same uncertainty in the value of the COBE
normalization, particularly when a parameter such as $n$ is varied. Thus,
normalizing to $\sigma_8$ rather than to COBE reduces any purely
normalization-related effects of small variations in $n$ on the amount of
power available on small scales, due to the shorter lever arm between the
physical scales associated with $\sigma_8$ and reionization. This is an
important consideration for this paper, one of whose results demonstrate
the sensitivity of $z_{\rm reion}$ to $n$.

In this work, we focus specifically on the constraints anticipated from the
data from the $MAP$ satellite. We include the effects of instrumental
noise, rather than assuming cosmic variance limited data as in Paper I. We
take experimental specifications and the method of constructing $F_{ij}$
from \citet{eht99}, and assume that foregrounds can be effectively
subtracted from $MAP$ data \citep{tehoc}.  In all the figures below, the
error ellipses, where displayed, represent 68\% joint confidence regions.

\section{Results}

We now discuss the constraints that a stellar reionization model and CMB
parameter forecasts may place on each other. We define our standard model
(SM) as described by the 7-parameter set, [$\sigma_8$, $\Omega_b$, $h$,
$n$, $\Omega_\Lambda$, $\Omega_{\rm M}$, $\tau$/$f_\star$] = [1.0, 0.04,
0.7, 1.0, 0.7, 0.3, 0.048/0.05]. As mentioned earlier, $\Omega_{\rm K}$ is
fixed to be [1 - $\Omega_{\rm M}$ - $\Omega_\Lambda$] in the parameter
analyses below, and its value is zero only in the SM.  We set the clumping
factor $c_L$ = 30 (see, e.g., \citet{mhrees99}, and references therein),
which, together with our choice of $f_{esc}$ = 0.1 (\S 2), leads to $\tau
\sim$ 0.048 for the SM, corresponding to a reionization epoch of $z_{\rm
reion} = 8$. The average ionization fraction of the IGM in the SM is
$10^{-4}$, $10^{-3}$, 0.01, 0.1, and 1.0 at the respective redshifts of
about 18, 15.6, 13.2, 10.6, and 8. This is consistent with the evolution of
the volume-averaged hydrogen ionization fraction in numerical simulations
of reionization with the same background cosmology as the SM here (see,
e.g., Figure 10 of Gnedin 2000.)

Our choice of parameters for the SM, though well-motivated and in
concordance with a variety of observations, is deliberately constructed to
generate late reionization, given the recent observational claim of
detecting the last stages of reionization at $ z \sim 6.3$. The
semi-analytic treatment here defines reionization as the overlap of H~II
regions, and corresponds to the component of the IGM that dominates the
ionization by volume filling factor at high redshift.  By this definition,
reionization somewhat precedes the disappearance of the GP trough in the
IGM \citep{hl99}, which represents the ionization of any remaining H~I in
highly overdense or clumped portions of the IGM, or in individual H~II
regions. Note also that the power of the CMB to constrain cosmological
parameters is often better demonstrated by considering parameter
combinations such as $\Omega_b h^2$ and $\Omega_{\rm M} h^2$ rather than
individual ones.  We have chosen our parameter as displayed above in order
to make an apples-to-apples comparison between the dependency of the CMB
power spectra and the reionization model on these individual cosmological
parameters. Furthermore, our intent in this work is to show that a
reionization model can tighten constraints from the CMB on, e.g., $n$, but
not on parameter combinations such as $\Omega_b h^2$ and $\Omega_{\rm M}
h^2$ because these will be very well-determined by CMB data alone.

As a reference, we show in Figure 1 the angular temperature and
polarization power spectra of the CMB for the SM. As we noted earlier, the
main effect of the reionization of the IGM is an overall damping of the
primary CMB temperature and polarization anisotropies. It also generates a
new feature in the CMB polarization spectrum corresponding to the horizon
size associated with $z_{\rm reion}$; for the late reionization in our SM,
this corresponds to the polarization bump at $l \la 5$. The signal
associated with this unique probe of reionization has an extremely small
value, being less than the temperature anisotropy by over two orders of
magnitude at these scales.

\begin{figure}[htb]
\begin{center}
\includegraphics[height=3.5in]{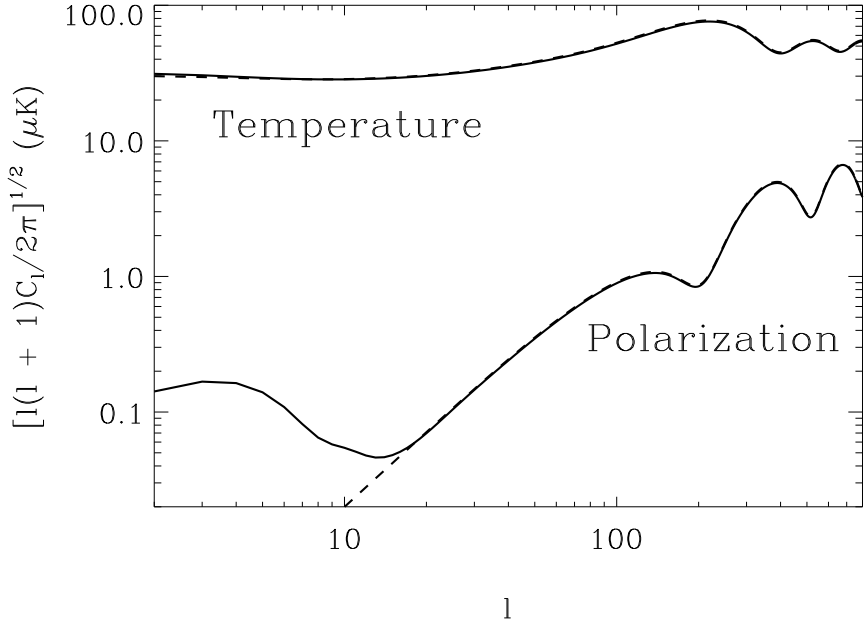}
\end{center}
\caption{Theoretical temperature and $E$-channel polarization angular power
spectra of the CMB in units of $\mu$K for this work's standard model (SM),
shown by the solid line: $\sigma_8$ = 1.0, $\Omega_b$ = 0.04, $h$ = 0.7,
$n$ = 1.0, $\Omega_\Lambda$ = 0.7, $\Omega_{\rm M}$ = 0.3, $\tau$ =
0.048. Dashed lines display the power spectra for the same choice of
cosmological parameters with $\tau$ = 0; the difference is noticeable only
for $l \la 20$.}
\end{figure}

\subsection{Using a Reionization Model to Improve Constraints from the CMB}

Using the techniques in \S 2, we can use a reionization model to constrain
cosmological parameters beyond the limits obtained from CMB data alone,
through $\tau$ or $f_\star$.  Let us first focus on the former
case. Certain combinations of parameters are well known to be degenerate in
CMB parameter extraction, such as $\tau$--$\sigma_8^2$ and $\tau$--$n$
(see, e.g., the recent analyses by the DASI, MAXIMA-1 and Boomerang
collaborations). A reionization model can provide complementary
information, as $\tau$ is itself a function of cosmological parameters, and
break such degeneracies. We display this in Figures 2 and 3, for the above
combinations of degenerate parameters, where we marginalize only over the
respective two-dimensional spaces and keep all the other parameters fixed
at their SM values. The dark outer and light inner ellipses correspond to
the 1 $\sigma$ constraint from $MAP$'s temperature (T), and temperature
plus polarization (T+P) data. The thin solid line represents the functional
dependence of $\tau$ on $n$ or $\sigma_8^2$ from the reionization model for
$f_\star$ = 0.05, and the dashed lines represent the possible range for
$\tau$, given the uncertainty in the value of $f_\star$. This possible
range for $f_\star$ of $\sim$ 0.01--0.15 comes from the results of
numerical simulations and from arguments of avoiding excessive metal
pollution of the IGM at late redshifts (Paper I, and references therein);
it represents the astrophysical uncertainty in our chosen reionization
model, given the choice to set those $P_{\rm cosmo}$ other than $n$ or
$\sigma_8^2$ to their values in the SM. Figures 2 and 3 show that the
reionization model can be valuable in breaking degeneracies in CMB
parameter analyses, even when given the range in the potential values of
$f_\star$.

\begin{figure}[t]
\begin{center}
\includegraphics[height=3.5in]{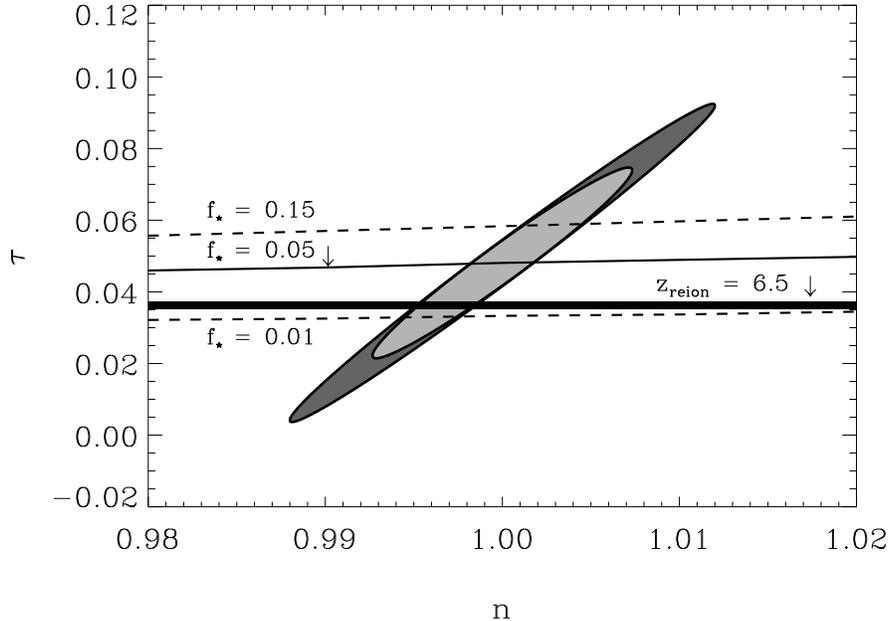}
\end{center}
\caption{Constraints from the reionization model and projected data from
$MAP$ in the $\tau$--$n$ plane, after 2-D marginalization over the [$\tau$,
$n$] space with all other parameters fixed at their SM values.  The thin
solid line displays $\tau$ as a function of $n$ from the reionization model
with $f_\star = 0.05$, and the dashed lines represent the astrophysical
uncertainty in $\tau$, given the permitted range of 0.01--0.15 in the value
of $f_\star$. The dark outer and light inner ellipses correspond to the 1
$\sigma$ joint confidence regions from $MAP$'s temperature, and temperature
plus polarization data. Note the strong dependence of $z_{\rm reion}$, and
hence $\tau$, on $n$ through the reionization model (thin solid line): for
$n$ = 0.98--1.02, $z_{\rm reion}$ $\sim$ 7.75--8.2.  The thick solid line
represents the constraint from a hypothetical independent measurement of
$z_{\rm reion}$ = 6.5.}
\end{figure}
\begin{figure}[t]
\begin{center}
\includegraphics[height=3.5in]{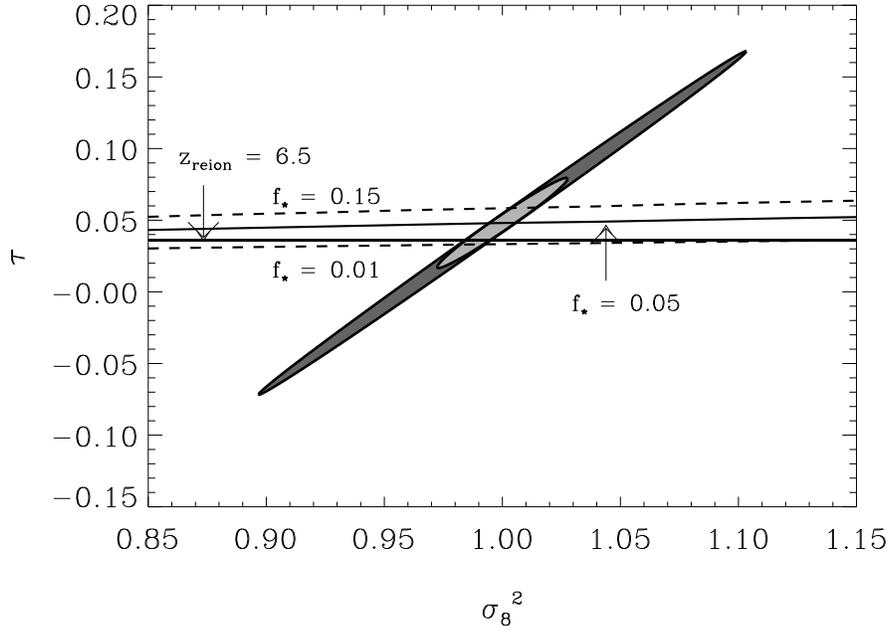}
\end{center}
\caption{Constraints from the reionization model and projected data from
$MAP$ in the $\tau$--$\sigma_8^2$ plane, after 2-D marginalization over the
[$\tau$, $\sigma_8^2$] space with all other parameters fixed at their SM
values. Plot legend is the same as in Figure 2.}
\end{figure}

The main source of the dependence of $\tau$ on $n$ and $\sigma_8^2$ is
$z_{\rm reion}$, and to a lesser extent, the term $f_\star F_{\rm B}$ in
eqn. 1, which is never more than a 2\% effect in the value of $\tau$ for
the SM.  Ideally, we would like to characterize $\tau$ as a function of
$P_{\rm cosmo}$, in order to eliminate its dependence on the astrophysical
details of reionization.  If we neglect the term $f_\star F_{\rm B}$,
eqn. 1 is considerably simplified as $F_{\rm H \, II}$ = 1.0 along the
line-of-sight from the present ($z = 0$) to $z = z_{\rm reion}$. The
problem now reduces to parametrizing $z_{\rm reion}$ in terms of $P_{\rm
cosmo}$ alone; in reality, however, $z_{\rm reion}$ is a non-unique
function of various cosmological parameters as well as the specific
(astrophysical) reionization scenario. The analysis of \citet{griffiths},
while having the advantage of being fitted to the available data at the
time, encountered the same problem of being unable to uniquely relate
$z_{\rm reion}$ to the cosmological parameters that they considered ($h$,
$n$, $\Omega_0$); the fit provided by them for $\tau$ as a function of
these three parameters was purely empirical but not based on any model of
the reionizing sources. Thus, the {\it only} way to utilize the valuable
sensitivity of $\tau$ to $n$ and $\sigma_8^2$ is via a reionization
model. The importance of retaining the information contained in $z_{\rm
reion}$, particularly for the lower bound on $n$, was noted in
\citet{covilyth}, where they pointed out that the choice to leave $z_{\rm
reion}$ as a free parameter, e.g., in the analysis of \citet{tzh01}, could
lead to an artificially lowered value of $n$ from CMB data.

\begin{figure}[htb]
\begin{center}
\includegraphics[height=3.5in]{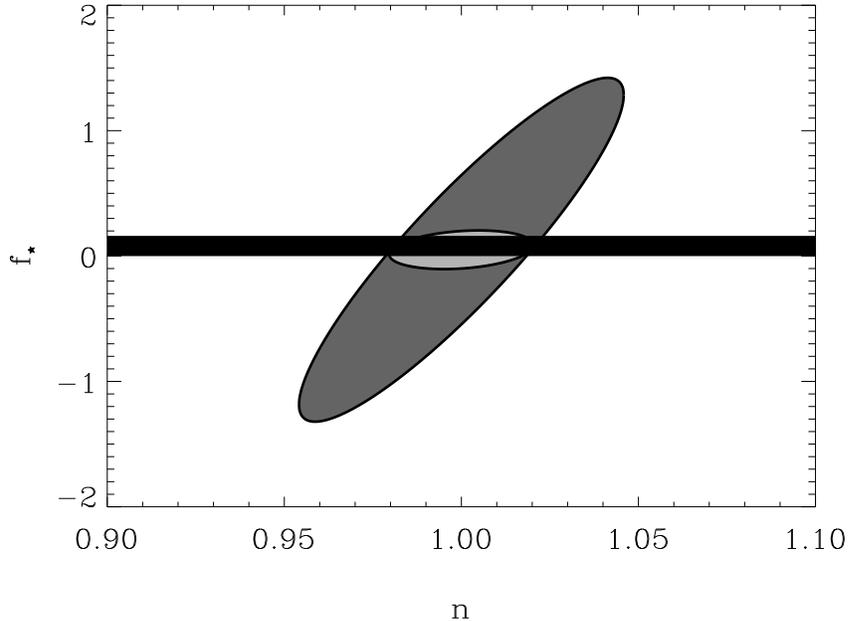}
\end{center}
\caption{Constraints from the projected data from $MAP$ in the
$f_\star$--$n$ plane after full 7-D marginalization over the [$f_\star$,
$P_{\rm cosmo}$] space. The dark outer and light inner ellipses correspond
to the 1 $\sigma$ joint confidence regions from $MAP$'s temperature, and
temperature plus polarization data. Solid horizontal band represents the
entire allowed astrophysical range of 0.01--0.15 for $f_\star$.}
\end{figure}

What if, however, there were an independent limit on $z_{\rm reion}$? One
possible method, which involves relating $z_{\rm reion}$ directly to the
fraction of baryons in star-forming halos, $F_{\rm B}$, has been explored
by \citet{covilyth} and \citet{tsb94}. Subject to theoretical
uncertainties, this is well motivated, as regardless of the details of the
nature and the sources of reionization, one requires in the end a certain
number of IGM-ionizing photons per baryon in collapsed structures. Another
possibility, which may shortly be upgraded to reality, would be a
spectroscopic detection of $z_{\rm reion}$ through the GP effect in the
absorption-line spectra of the highest-$z$ sources (see \S 1). The great
advantage of this second kind of independent determination of $z_{\rm
reion}$ is that one may safely bid farewell to the pesky details of
``gastrophysics'' \citep{bond} in parametrizing $\tau$ for CMB parameter
extraction! If we drop the term $f_\star F_{\rm B}$ in eqn. 1, we can now
relate $\tau$ to the $P_{\rm cosmo}$ other than $n$ and $\sigma_8^2$. This
leads to {\it a unique value of $\tau$ in the 2-D space of Figures 2 and
3}, which is depicted as the thick solid line for a hypothetical measured
value of 6.5 for $z_{\rm reion}$. Such a detection can be useful in
breaking parameter degeneracies and improving constraints from the CMB
without invoking a specific reionization model.  Note that a detection of
$z_{\rm reion}$ cannot be translated to a unique prior on $\tau$ for
multi-parameter marginalization, as the latter is also determined by
cosmological parameters such as $\Omega_\Lambda$, $\Omega_b$, etc.  Thus,
an independent determination of $z_{\rm reion}$ is best utilized in the 2-D
spaces of parameter combinations that are degenerate with $\tau$, such as
the examples in Figures 2 and 3.

We now move on to the second case defined at the beginning of this section,
where one may translate astrophysical limits to constrain cosmology. We
marginalize over the 7-D space of [$f_\star$, $P_{\rm cosmo}$] rather than
[$\tau$, $P_{\rm cosmo}$], by using the reionization model to relate them
via $\tau$ (eqns. 2 and 3). We can then apply independent limits on
$f_\star$ (0.01--0.15) to further constrain $P_{\rm cosmo}$. Figure 4
displays one such case in the $f_\star$--$n$ subspace for the projected
constraints from $MAP$'s T and T+P data. Despite the error ellipses being
lower bounds to those that $MAP$ will provide (given our assumption of
successful foreground removal), the {\it entire} astrophysical permitted
band for $f_\star$ can still reduce the 1-$\sigma$ error for $n$. Although
one may propose alternate ranges for $f_\star$, we anticipate that the main
point here-- that known constraints on $f_\star$ have the power to
strengthen limits from the CMB on $P_{\rm cosmo}$-- will still hold true.

In summary, using a reionization model can break degeneracies in CMB
parameter estimation related to $\tau$, and improve the errors from $MAP$
data on $n$ and $\sigma_8^2$ by factors of at least 3--6 and 3--10
respectively for the case of $f_\star = 0.05$.  Alternatively, known
astrophysical limits on $f_\star$ can reduce the errors on $P_{\rm cosmo}$
from $MAP$, e.g., by up to a factor of 2 for $n$ from $MAP$ temperature
data. The strongest cross-constraint in the near future may be provided by
an independent measurement of $z_{\rm reion}$, which could reduce the 1
$\sigma$ errors on parameters that are degenerate with $\tau$, such as $n$
or $\sigma_8^2$, by factors of 3--10. The non-trivial advantage of this
last method is that it is independent of one's choice of reionization
model.

\subsection{Using CMB Data to Constrain a Reionization Model}

Given the framework of this paper, there are at least two ways that
forthcoming CMB data may be used to constrain aspects of reionization.
First, we can use a reionization model to extract [$f_\star$, $P_{\rm
cosmo}$] rather than [$\tau$, $P_{\rm cosmo}$] from CMB data.  Table 1
displays the 1 $\sigma$ errors from $MAP$ T and T+P data for full
marginalization over the 6-D [$P_{\rm cosmo}$] and the 7-D [$\tau$, $P_{\rm
cosmo}$] parameter spaces. Both the 6-D and 7-D cases assume 65\% sky
coverage and factor in the effects of instrumental noise for $MAP$.
Including $\tau$ in the analysis significantly worsens error bars from
$MAP$'s T-data, particularly for $\sigma_8^2$ and $n$; this can be expected
from the degeneracies discussed above. Put another way, excluding $\tau$ or
setting it to be zero can lead to deceptively small errors in parameters
such as $\sigma_8^2$ and $n$.

\begin{table}[t]
\renewcommand{\arraystretch}{1.2}
\begin{center}
\caption{Projected 1 $\sigma$ errors from $MAP$ data}
\vspace{0.4in}
\begin{tabular}{lc|cc|cc|c}
\tableline \tableline
 &  & \multicolumn{2}{c|}{Without $\tau$} & \multicolumn{2}{c|}{With $\tau$} & Ideal $MAP$   \\ \tableline
 & & &  &  &  & \\ 
Parameter & & T & T+P & T & T+P &  T+P  \\ \tableline
$\tau$ [$f_\star$] &  & &  & 0.193 [0.904]  & 0.022 [0.102] & [0.011] \\ 
$\sigma_8^2$ & & 0.048 & 0.047 & 0.079 & 0.047 & 0.026 \\
$\Omega_b$ & & 0.003 & 0.003 & 0.004 & 0.003 & 0.002 \\
$h$ & & 0.022 & 0.022 & 0.026 & 0.022 & 0.012  \\
$n$ & & 0.013 & 0.013 & 0.03 & 0.013 & 0.008 \\
$\Omega_\Lambda$ & & 0.018 & 0.018 & 0.022 & 0.018 & 0.008 \\
$\Omega_{\rm M}$ & & 0.019 & 0.018 & 0.019 & 0.018 & 0.01 \\

\tableline
\end{tabular}

\tablecomments{The 1 $\sigma$ errors anticipated from $MAP$, with
temperature (T), and temperature plus polarization (T+P) data. The
respective columns are-- Without $\tau$: 6-D marginalization over [$P_{\rm
cosmo}$] space only; With $\tau$: 7-D marginalization over the full
[$\tau$, $P_{\rm cosmo}$] space; Ideal $MAP$: sky coverage of 50\%, and
data is cosmic variance limited to $l \sim 500$. With/without $\tau$
columns assume 65\% sky coverage and include the effects of instrumental
noise for $MAP$.  Note that excluding $\tau$ from the analysis leads to
deceptively small errors for $n$ and $\sigma_8^2$ from the temperature
data. Entries in brackets represent 1 $\sigma$ errors from 7-D
marginalization over [$f_\star$, $P_{\rm cosmo}$] space, using the
reionization model.}

\end{center}
\end{table}

Using the reionization model now to relate $f_\star$ and $P_{\rm cosmo}$
(eqns. 2 and 3), we see from Table 1 that $MAP$'s T and T+P data do not
constrain $f_\star$ very strongly.  If, however, $MAP$ can achieve being
cosmic variance limited to $l \sim 500$ with 50\% sky coverage, which we
label as ``Ideal $MAP$'' in the table, it is possible with T+P data to
determine $f_\star$ to significantly greater accuracy than its currently
allowed range. Given that we have not factored in foreground contamination
of the CMB polarization signal, which particularly degrades parameter
extraction on the (large) scales at which reionization has a unique
signature \citep{tehoc,bacci}, our prediction of strong limits on $f_\star$
from the CMB may be somewhat optimistic.

A second possibility involves using a measurement of $\tau$, particularly
through polarization in the CMB; current data place only rough upper limits
of $\tau$ $\la$ 0.3. A low net value of $\tau$ would imply that star
formation cannot be widespread, or that it has to be fairly
inefficient. The majority of the theoretical models to date imply that
reionization takes place between $z \sim$ 8--20. In our SM, we allow star
formation only in halos of virial temperature $\ga$ 10$^4$ K. If we replace
this mass threshold scale in our Press-Schechter evolution with the Jeans
mass scale at each redshift, then, for the SM cosmological parameters, we
obtain $\tau \sim 0.078$ (0.11), and $z_{\rm reion} \sim 11.25$ (14.2) with
(without) clumping. Thus, as an example, if $\tau$ were measured in the
future to be $\la$ 0.05, it would imply that early star formation has to be
relatively rare, i.e., occurs in high-mass rather than in low-mass halos at
high redshifts, or that it is relatively inefficient (values of $f_\star$
significantly less than in the SM). While this statement relies on our
assumptions and adopted reionization model in this work, it is a potential
constraint in the near future.

In summary, forthcoming CMB data may be able to constrain the fraction of
baryons that participated in early star formation, and, more speculatively,
the sites of such stellar activity as well, if reionization were caused by
stars.

\section{Discussion}

In this work, we have focussed primarily on tightening constraints on
cosmological parameters and on constraining aspects of the reionizing
source population from CMB data by using either a semi-analytic
reionization model for $\tau$ or an independent determination of $z_{\rm
reion}$. For this purpose, we have considered only the first-order effects
of reionization: the damping of the primary CMB anisotropies and the
generation of a new polarization signal at large angular scales. This was
motivated by the possibility of these signatures being detected imminently,
and by the simplicity of parametrizing them. The reader may, however,
wonder to what degree the results presented here depend on the adopted
reionization model, and on the effects of inhomogeneous reionization and
the secondary CMB anisotropies generated during reionization which were not
considered here. We address these issues below; we also discuss which
observations of the CMB and high-$z$ large scale structure have the
potential to constrain not only $z_{\rm reion}$, but also the duration,
inhomogeneity and physical processes related to reionization.

We have chosen stars as the principal reionizing source population for the
observationally motivated reasons described in \S 1. We emphasize that
other possibilities, such as a large high-$z$ population of mini-quasars
that are as yet undetected, or a combination of stars, quasars and other
sources cannot be ruled out at present. The methods in this paper are,
however, easily extended to other reionization scenarios. Since the
relevant quantity for reionization is the average ionizing efficiency of
each baryon in luminous objects, the constraints in this paper involving
$f_\star$ may be equivalently regarded as limits on the fraction of baryons
in individual halos that formed objects with a star-like (soft) ionizing
spectrum.  The calculations in this paper can be applied equally well to
sources with QSO-like (hard) ionizing spectra, including the case of
metal-free stars, but such reionization models face the added consideration
of accounting for the observed lag of the He~II reionization epoch relative
to that of H~I.

Although we include the effects of IGM clumping in this paper, the
development of luminous objects and the gradual overlap of H~II regions are
themselves characterized only in an average homogeneous sense. In reality,
the first astrophysical sources of ionizing photons are likely to be
located in very dense regions embedded in the large scale filamentary
structure of matter, so that reionization is a highly nonlinear,
inhomogeneous process.  To probe the complex details of this so-called
patchy reionization, one must turn to numerical simulations, which can
follow the detailed radiative transfer and reveal the full 3D topology of
reionization. Simulations are also very useful for quantifying the
secondary anisotropies generated in the CMB during reionization, because
they can perform the necessary characterization of the spatial variation of
the ionization levels. Inhomogeneous reionization generates second-order
temperature anisotropies in the CMB, with contributions from the spatially
varying electron density and the bulk velocity field of the electrons. The
first effect can be caused by variations in the baryon density
\citep{ostvishniac} or in the ionization fraction (patchy reionization),
the latter depending on the typical size of H~II regions around individual
sources and on the spatial correlations of ionized gas.

The amplitude of these second-order features can in principle constrain
$z_{\rm reion}$, as well as the nature and sites of the reionizing sources
through the angular scale on which they generate a secondary signal in the
CMB.  Detailed studies of these effects (see, e.g., \citet{gnedinjaffe},
and references therein, \citet{benson}) indicate, however, that the
secondary anisotropy spectrum is dominated by the thermal
Sunyaev--Zeldovich effect from low-$z$ clusters of galaxies at CMB
multipoles $l \sim$ 1000--$10^5$, with the kinetic Sunyaev--Zeldovich
effect taking over at $l \ga 10^5$.  The signal from the patchiness of
reionization (the spatial variation of the ionization fraction) appears to
be subdominant at all angular scales to the two effects above, if the
ionizing sources are clustered on the scales that are typical of stellar
and mini-QSO models. The signature of patchy reionization in the CMB may be
detected, however, if reionization were caused by spatially rare, bright
objects having comparatively large H~II regions, although this will still
occur at extremely large $l$, and the sources in such a case cannot be
large bright QSOs from the discussion in \S 1. The geometry of reionization
also partly determines the amplitude of such secondary anisotropies; the
signal is relatively low if the low-density IGM regions are ionized first,
which is likely. For the signal from patchy reionization to be dominant,
$z_{\rm reion}$ would have to exceed its current upper bound of $\sim$ 30
from CMB data. Lastly, the dominant contributions to secondary CMB
anisotropies from reionization come from the epochs just preceding $z_{\rm
reion}$, and from the correlations between high-density regions, which
trace the underlying matter density field rather than the varying
ionization of the IGM itself \citep{valsilk2}.

Thus, as \cite{gnedinjaffe} state, the prospects for describing the details
of reionization, such as its inhomogeneity and duration, through secondary
anisotropy signatures at sub-degree scales in the CMB are not very
encouraging, as they are not likely to be within the detection capabilities
of the next decade of CMB experiments.  Although numerical simulations,
given their computational expense, are not a practical method of
quantifying the role of patchy reionization in the likelihood analyses in
CMB parameter estimation, their findings indicate that second-order effects
from reionization are not likely to contaminate parameter extraction from
CMB data in the near future.  This justifies our approach in this work
where we have concentrated on the first-order effects of reionization on
the CMB; most of the information on $\tau$ comes from a large-scale
polarization signal, which is not complicated by the inhomogeneity of
reionization manifesting in the CMB at much smaller angular scales.

The semi-analytic reionization model used in this paper is formulated
through a generally accepted prescription with simple physics, thereby
reducing a large body of possible input parameters to the essential ones.
The advantage of using a model for $\tau$ is that we have a simple
well-motivated model that does not introduce additional parameters (unlike
Paper I), and that relates the CMB data to the most important quantity
driving reionization. In the case of the reionization model considered
here, this quantity happens to be $f_\star$, but it could equivalently be
$f_{esc}$ or other parameters that describe the sources of reionization.
We have also demonstrated how future CMB data may be able to constrain the
sites of early star formation, as well as the fraction of baryons that
participated in it, assuming a stellar reionization scenario. Lastly, a
reionization model allows us to strongly constrain parameters such as $n$
with CMB data, as $z_{\rm reion}$ is very sensitive to the amount of
small-scale power.

The drawback of some of the results presented here is the use of
model-dependent means to break degeneracies or tighten constraints on
cosmological parameters in CMB data analyses. It would be more preferable
to combine data sets to accomplish this, which numerous papers have
explored using, e.g., weak lensing data or large scale structure data from
the {\it IRAS} PSCz Survey, the Sloan Digital Sky Survey (SDSS), the 2dF
Survey, etc. (\S 1). Although such data of the local universe will not
directly constrain $\tau$, they will add information that is complementary
to the CMB on parameters such as $n$ and the power spectrum normalization,
thereby helping to break degeneracies between these parameters and $\tau$
in CMB parameter estimation. Hence, the use of a reionization model may not
be necessary.  One may then ask what sort of data related to reionization
will prove valuable for CMB analyses.  We have taken a first step in this
work towards answering this question by demonstrating the power of an
independent determination of the reionization epoch (Figures 2 and 3) to
break parameter degeneracies in the CMB, and to sharply reduce the error on
parameters such as $n$.

The most promising data avenue to probe the sources, duration, and
inhomogenous aspects of reionization will likely be high-resolution
spectroscopic studies with high signal-to-noise of bright quasars and
starforming galaxies at $z \ga$ 6. The current evidence for a complete H~I
GP trough, and hence $z_{\rm reion}$, comes from the spectrum of a single
$z \sim$ 6.3 QSO. The acquisition of more data along many more lines of
sight to sources at $z \sim$ 6--10 is required to adequately represent how
the appearance and duration of the GP trough varies with redshift and
different sightlines through the IGM. This will lead to an angular map on
the sky of $z_{\rm reion}$ as a function of sightline, which will
dramatically increase our power to quantify the spatial nonuniformity of
the reionization process, the size distribution of ionized regions, the
nature of the ionizing sources, and the physical conditions in an average
region of the IGM during reionization \citep{barkana1}.  Such observations
are within the capabilities of the SDSS, which should detect about 20
bright quasars at $z \ga$ 6 during the course of the survey \citep{becker},
and are important targets in the planning of the {\it Next Generation Space
Telescope}.  Such data would also permit the direct extraction of $z_{\rm
reion}$ from the portions of transmitted flux between the individual
troughs from the Lyman series lines for sources that lie just beyond
$z_{\rm reion}$ \citep{hl99}, although this could be complicated by
intervening IGM clumpiness or damped Ly$\alpha$ systems. Such spectroscopic
studies will be invaluable for characterizing the scale dependence of the
high-$z$ IGM's porosity in the epochs around $z_{\rm reion}$; the challenge
will be to understand which of the cosmic variance in the data arises from
the underlying mass distribution rather than the ``gastrophysics''
manifested through patchy ionization.

In addition to Ly$\alpha$ GP trough studies, \citet{umemura} have suggested
that the H$\alpha$ forest is a significantly more powerful probe of the
ionization history of the universe at $z \ga 5$. The H$\alpha$ line is more
sensitive to small changes in the degree of ionization when the IGM neutral
fraction is at levels of 1\% or below, unlike the resonant Ly$\alpha$ line
which can cause regions of the IGM with even a small amount of H~I to
transmit close to zero flux. Thus the H$\alpha$ forest may provide better
constraints of the IGM during the epochs spanning reionization.  Lastly,
the neutral IGM prior to complete reionization may be detected in 21-cm
emission or absorption against the CMB, depending on whether the IGM
experienced any heating in association with reionization.  Future radio
telescopes can perform this tomography, constraining the thermal and
density properties of the pre-reionization IGM, as well as the distribution
of H~II regions \citep{tozzi}. Such observations, however, are subject to
the same problem as those of CMB polarization, which is the effective
removal of galactic and extragalactic foregrounds. Other observational
probes of the epoch and sources of reionization may be found in
\citet{hk99}, and \citet{loebbark}.

In summary, the near-term prospects for determining the epoch of
reionization from data of the CMB and of high-$z$ large scale structure
appear excellent. Although it is possible to constrain the astrophysical
aspects of reionization by combining a reionization model with imminent CMB
data, limits on the sources, duration and inhomogeneity of reionization
from data alone are likely to take at least several years.

\section{Conclusions}

We have extended previous work on the mutual constraints that are possible
between a reionization model and parameter estimation from CMB data to a
more general parameter set in a $\Lambda$CDM cosmology, and for the data
anticipated from the $MAP$ satellite. A reionization model provides
valuable complementary information for cosmological parameter extraction
from the CMB. In particular, the well-known $\tau$--$\sigma_8^2$ and
$\tau$-$n$ degeneracies, which continue to be present in the most recent
data from the DASI, MAXIMA-1 and Boomerang experiments, can be broken (see
Figures 2 and 3), even when allowing for the effects of the astrophysical
uncertainty in the reionization model for $\tau$. Furthermore, using the
reionization model in this work improved the projected errors on $n$ and
$\sigma_8^2$ from $MAP$ data by respective factors of about 3--6 and 3--10.

Alternatively, we may use the reionization model to relate the astrophysics
of reionization to cosmology: independent theoretical limits on $f_\star$
can reduce the forecasted errors on $P_{\rm cosmo}$ from $MAP$, e.g., by up
to a factor of 2 for $n$ (Figure 4). Applying reionization models to CMB
data provides the only way, in the absence of an alternate determination of
$z_{\rm reion}$, to utilize the strong sensitivity of $\tau$ through
$z_{\rm reion}$ to parameters such as $n$ and $\sigma_8^2$, which are
important inputs to models of inflation and the evolution of structure.
The specific dependence of $z_{\rm reion}$ on $n$ through the reionization
model can be seen in Figure 2: for the $f_\star = 0.05$ case, $z_{\rm
reion}$ increases from 7.75 to 8.2, with respective values of $\tau$ from
$\sim$ 0.046 to $\sim$ 0.05, as $n$ varies from 0.98 to 1.02.

Forthcoming CMB data also have the potential to constrain the sites of
early star formation, as well as the fraction of baryons that participate
in it, if reionization were caused by stellar activity at high redshifts
(\S 3.2).  In particular, if $MAP$ can achieve 50\% sky coverage and is
cosmic variance limited to $l \sim 500$, the 1 $\sigma$ error for $f_\star$
could be significantly smaller than the current uncertainty in its value
(Table 1, ``Ideal $MAP$'' column), although it requires a detection of
polarization in the CMB at large angular scales. This polarization signal
is, however, of sufficiently small magnitude for late reionization (Figure
1) that it will prove extremely challenging to detect experimentally,
especially when foregrounds are included, which we have assumed here can be
effectively subtracted.  Thus, the utility of CMB data in constraining the
astrophysical aspects of reionization, besides being model-dependent, is
optimistic at best.

While the anticipated errors from $MAP$ in Table 1 are dependent on the
size of our chosen parameter space, any analysis of CMB data cannot include
very many fewer parameters than we have considered here. Larger parameter
spaces and the inclusion of foregrounds will only increase the projected
errors in this work, thereby enhancing the importance of techniques to
break parameter degeneracies, including the three presented here-- the use
of a reionization model, applying known astrophysical limits, and an
independent measurement of the reionization epoch. The last method appears
particularly promising from the recent detection of a GP trough in the
spectrum of a quasar at $z \sim 6.3$ \citep{becker}, which could well
represent the last stages of non-uniform reionization.  We anticipate that
this may provide the strongest cross-constraint in the near future, which
we have shown (\S 3.1) could reduce the 1 $\sigma$ errors on parameters
that are degenerate with $\tau$, such as $n$ or $\sigma_8^2$, by factors of
3--10 for data from $MAP$. The great advantage of using a detection of
$z_{\rm reion}$ to break such degeneracies is that it is not subject to the
details of ``gastrophysics'' that partly determine the optical depth to
reionization.  A measurement of $z_{\rm reion}$ cannot necessarily be
translated to a unique prior on $\tau$ in multidimensional analyses, as the
latter is also determined by cosmological parameters.  Thus, an independent
determination of $z_{\rm reion}$ is best utilized in the specific parameter
spaces that are degenerate with $\tau$ (Figures 2 and 3).

In conclusion, this is a special time for cosmology (and for those employed
in its study!), when observational efforts to detect the epoch of hydrogen
reionization are rapidly narrowing the bracketed range of possible
redshifts-- from the lower end, through spectroscopic studies of the
highest-redshift objects, and from the upper end, with data from past and
ongoing CMB experiments. This has provided a unique opportunity to jointly
test theoretical models of the CMB and of the growth of structure, in order
to understand the nature and birth sites of the first luminous objects. We
can look forward to the next few years of data from such endeavors, which
are likely to settle important frontiers in cosmology including the epoch
when the universe returned to a fully ionized state.

\acknowledgements

We thank Andrew Hamilton, Mark Giroux, Kim Coble and Nick Gnedin for
helpful comments, Matias Zaldarriaga and Daniel Eisenstein for useful
correspondence, and Rocky Kolb for past conversations in which we first
heard of such choice terms as ``gastrophysics''. We gratefully acknowledge
support from NASA LTSA grant NAG5-7262.

\end{document}